# Dynamics of person-to-person interactions from distributed RFID sensor networks

Ciro Cattuto[a,*], Wouter Van den Broeck [a] , Alain Barrat[b,a], Vittoria Colizza[a], Jean-François Pinton[c], Alessandro Vespignani[d,e,f]

a) Complex Networks and Systems Group, Institute for Scientific Interchange Foundation, Turin, Italy
b) Centre de Physique Théorique (CNRS UMR 6207), Marseille, France
c) Laboratoire de Physique de l'ENS Lyon (CNRS UMR 5672), Lyon, France
d) Center for Complex Networks and Systems Research, School of Informatics and Computing, Indiana University, Bloomington, IN, USA
e) Pervasive Technology Institute, Indiana University, Bloomington, IN, USA
f) Institute for Scientific Interchange Foundation, Turin, Italy

* Corresponding author: ciro.cattuto@isi.it

## Abstract

Digital networks, mobile devices, and the possibility of mining the ever-increasing amount of digital traces that we leave behind in our daily activities are changing the way we can approach the study of human and social interactions. Large-scale datasets, however, are mostly available for collective and statistical behaviors, at coarse granularities, while high-resolution data on person-to-person interactions are generally limited to relatively small groups of individuals.  Here we present a scalable experimental framework for gathering real-time data resolving face-to-face social interactions with tunable spatial and temporal granularities.

We use active Radio Frequency Identification (RFID) devices that assess mutual proximity in a distributed fashion by exchanging low-power radio packets. We analyze the dynamics of person-to-person interaction networks obtained in three high-resolution experiments carried out at different orders of magnitude in community size. The data sets exhibit common statistical properties and lack of a characteristic time scale from 20 seconds to several hours. The association between the number of connections and their duration shows an interesting super-linear behavior, which indicates the possibility of defining super-connectors both in the number and intensity of connections.

These results could impact our understanding of all phenomena driven by face-to-face interactions, such as the spreading of transmissible infectious diseases and information.

## Introduction

Social sciences are being transformed by the possibility of collecting and analyzing the massive amount of digital information we leave behind in our daily activities [1-5]. This new opportunity, sometimes referred to as "reality mining" [6], provides insights into patterns of human life such as population flows inside cities, daily mobility patterns, or the geographical proximity of our social relations [7,8]. Along with these new empirical datasets, computational social science is emerging as a new way to study and predict social behavior [8]. One of the main issues in this context is the trade-off between the granularity of the data and the amount of information on each single interaction. In founding more sophisticated computational frameworks, it is of key importance to bridge the gap between scales and achieve a multi-scale view of social interactions. Experimentally, this calls for a scalable framework where the spatio-temporal resolution can be tuned and used to simultaneously probe different interaction scales, from co-presence in a room, to loose spatial proximity, down to face-to-face proximity of individuals. The aim is to reconcile the fine spatiotemporal evolution of the social network with the coarse-grained structure used at the large-scale population level.

At present, several techniques and methods are segmented in spatial and/or temporal resolution. Bluetooth and Wi-Fi networks allow the collection of data on specific structural and temporal aspects of social interaction patterns [9-12]. However, the spatial resolution achieved by these techniques is at best of the order of a few meters and, in general, spatial proximity or co-location of wearable devices are not necessarily a good proxy for a social interaction between the individuals carrying them. Monitoring human mobility and social relationships using mobile phone traces [7, 13-17] scales up to millions of individuals but provides no direct information on face-to-face interactions unless custom software is deployed. On the other side of the spectrum, the MIT Reality Mining project [10,11,18] collected rich multi-channel data on face-to-face interactions at the expense of deploying sophisticated "sociometric badges". Finally, systems based on image and video processing [19] provide the richest dataset but are computationally complex, require line-of-sight access to the monitored spaces, and can hardly cope with the unsupervised detection of face-to-face interactions and with large scales.

Here we report on a framework for monitoring social interactions that reconciles scalability and resolution using a sensing tier that consists of inexpensive and unobtrusive active RFID devices. The devices are capable of sensing face-to-face interactions of individuals as well as spatial proximity over different length scales down to one meter or less. The data collection and processing tiers allow tuning of the scale over which the interaction mapping works. The approach is highly scalable: We provide data from deployments at social gatherings involving from 25 to 575 individuals. Analysis of the results shows a remarkable self-similarity in the statistical signature characterizing personal interactions, despite the different social contexts and scales of the deployments. We also identify the general presence of super-connecting behaviour of highly interacting individuals, whose general interaction time increases non-linearly with the number of interactions. These features may play a crucial role in the study of dynamical processes over time-dependent networks of human contact, such as computational models of social and biological contagion, and the development of algorithms for mobile applications and wearable devices.

## Results and Discussion

Our strategy hinges on keeping the interaction resolution as the focal point of our experimental framework. We trade the possibility of acquiring extra information on person-to-person interactions (such as audio information) with the possibility of deploying a sensor network of unobtrusive devices that can scale up to thousands of people. To this aim, we have developed a sensing tier made of active RFID tags that can be embedded in a conference badge. These tags feature a bi-directional radio interface and transmit packets carrying a unique identifier and a data payload. Radio packets can be received by a system of readers installed in the environment, as well as by other tags located nearby. The exchange of low-power radio packets between tags can be used to measure tag proximity and to detect face-to-face interactions between individuals. We operate the system of RFID tags as a single distributed sensing network. Tags do not act as simple isolated beacons, broadcasting packet to a central infrastructure. Rather, they exchange low-power packets in a peer-to-peer fashion, to sense their spatial neighbourhood and assess proximity with other tags (Figure 1A). The proximity information collected locally is then uploaded to the reading infrastructure and relayed to a data collection system. Some important properties of this infrastructure need to be emphasized. First, proximity-sensing packets are emitted at several discrete power levels. Range is controlled both at the transmitting and receiving end, and can be tuned to detect tag proximity within 1-2 meters. The weakest power levels are used to assess face-to-face proximity as packets can only be detected when persons are facing one another (no shielding of the RFID tags) within about 1m. Moreover, RFID tags report about neighbouring devices using high power levels, thus few readers suffice to cover large indoor spaces. Finally, we operate the entire data collection pipeline in real time, enabling interactive applications and reflexive individual or social usage of the information we gather [20]. More details on the experimental setup and on the visualization are provided in the Materials and Methods section.

We deployed our contact-sensing platform in a number of different environments, presented in Figure 1. The deployments involved a number of participants ranging from 25 to 575 individuals. In our measurements the mechanisms of proximity detection are intrinsically statistical (see the setup description in the Materials and Methods section). We coarse-grain time in intervals of 20 seconds, over which we can assess proximity – or lack thereof – with high confidence, while maintaining a temporal resolution which is more than adequate to describe the fastest social interactions in a social gathering. The (tunable) spatial resolution is determined by the range over which tags can exchange low-power packets. For the *ISI* and *SFHH* deployments the devices were configured for a shortest spatial range of about 1 meter (for packets emitted at the lowest power), which affords the detection of face-to-face proximity. For the *25C3* deployment, the proximity detection range extended to 4-5 meters and packet exchange between devices was not necessarily linked to face-to-face presence, but rather reflected group structures in space that mix face-to-face interactions with looser casual proximity. This leads to a larger proportion of detected groups of three or four individuals, with respect to the number of pairs, in this specific deployment, as shown in Figure S1 through the comparison with the *SFHH* case.

The dynamical properties of these interaction patterns can be inferred by measuring the duration of contacts and the duration of the intervals between consecutive contacts [9,11,14,21]. We define the duration of a person-to-person contact consistently with the temporal coarse-graining described above: we consider two persons to be "in contact" during an interval of 20s if and only if their RFID devices have exchanged at least one packet at the lowest power level during that interval. After a contact has been established, it is considered ongoing as long as the devices continue to exchange at least one such packet for every subsequent 20s interval. Conversely, a contact is considered broken if a 20s interval elapses with no low-power packets exchanged. Figure 2A reports, for every deployment, the probability distribution of the durations of person-to-person contacts. The distribution displays large fluctuations, indicating that there are comparatively few long-lasting contacts and a multitude of brief contacts. Qualitatively, this behavior is not unexpected, and a similar result has indeed been reported for the duration of contacts between Bluetooth devices [10,14,21]. A striking feature exposed here is the similarity of the probability distributions for face-to-face interactions at close distance (*ISI* and *SFHH*) to the distribution observed for larger detection range (*25C3*). The spatial scale of the interactions is not a discriminating signature of the observed dynamical behavior (see also Figure S2).

Along with the duration of face-to-face contacts it is possible to track the dynamics of richer structures that bear relevance to the dynamical processes that can occur on the network of contacts: for example triadic interactions. A triangle involving individuals A, B and C is recorded when, within the same 20-second interval, packets are exchanged separately between each of the pairs A-B, A-C and B-C. A triangle breaks whenever any of the involved contacts break, hence we define the duration interval of a triangle in the same way as we did for pair-wise contacts. Panel B of Figure 2 reports the probability distribution for the duration of triangles. All measurements yield broad distributions, with the *25C3* case showing the longest tail, as triangles are more likely to be detected because of the longer detection range. It is also especially insightful to measure the duration of the interval between two consecutive contacts of a given individual with two distinct persons. In other words, if A starts a contact with B at time $t_{AB}$ , and then starts a different contact with C at $t_{AC}$, the inter-contact interval is defined as $t_{AC}$ - $t_{AB}$. Measuring this quantity is relevant for the study of causal processes (concurrency) that can occur on the dynamical contact network, such as for example information diffusion or epidemic spreading. The inter-contact intervals determine the timescale after which an individual receiving some information or disease is able to propagate it to another individual. Thus, the interplay between this timescale and the typical timescales of the spreading processes is crucial to diffusion processes. The probability distributions of inter-contact events show a broad tail across the three deployments, signaling the absence of a characteristic timescale (see panel C of Figure 2). Strikingly, and in contrast with the distributions of pair-wise contact durations, these distributions expose differences between deployments. In particular, the distribution of inter-contact intervals turns out to be broader when short detection ranges are considered (*ISI* and *SFHH*). In the context of spreading processes this would imply that various ranges of possible contamination would correspond to different distributions of times between successive spreading events.

The combination of high resolution and scalability we achieved allows us to address the crucial problem of the robustness of the observed distributions. In Figures 3A, S2, S3, we show that the same distributions are obtained not only across deployments, but also within a single deployment, across different intervals of time (from a few hours, to one full day, to the event as a whole). Figure 3A also displays distributions of contact durations of individual tags, showing that the observed heterogeneity of contact durations in the population is present also at the individual level.

Moreover, in experiments involving the tracking of individuals'

behavior, technical difficulties and human factors can both act as sources of data incompleteness. Participation is voluntary and not all individuals agree to have their contacts tracked.  People who agree to participate can still lose their badge, forget to wear it, wear it improperly, or tamper with the RFID tag by damaging it, shielding its antenna or removing its power source. Proximity relations can be detected only when they involve properly tagged individuals. Furthermore, data loss can occur because of technical failures in the data collection pathway, from the readers to the networking and computing infrastructure (see for example in Figure 1 the drop in the *25C3* timeline at the midnight of the 29[th]). In running the experiments, we deal with the above issues using a set of data quality flags and heuristics that allow us to spot problems and react promptly. Data incompleteness, nevertheless, is inherently unavoidable. In order to test the robustness of the data with respect to sampling and failure issues, we have simulated heavy data incompleteness by removing from the dataset the contacts involving a specified fraction of RFID tags, chosen at random. On the resulting decimated data set we have recomputed the distributions of the durations of contacts and triangles. Figures 3B and S4 show that these probability distributions are extremely robust with respect to the above sampling procedure: the shape is unchanged, and only the cutoff of the distribution moves to smaller values. These tests demonstrate that the behavior of the statistical distributions we measured is not altered by unbiased sampling of individuals, or by random data losses that may occur during the measurements. On the other hand, we cannot completely rule out that a systematic bias is introduced by the selection of volunteers, if volunteers and non-volunteers have different behavioral patterns. Accurately checking this point would require monitoring an independent data source for face-to-face contacts, and because of scalability issues this would be feasible only for small control groups.

Other biases may arise from our choice of deployment scenarios, as both of the large-scale deployment were conference-like gatherings. We recently collected data in radically different settings, namely a hospital, a school and a museum. The corresponding data analysis, which will allow to unveil similarities and differences across different scenarios, is currently work in progress.

The collected data afford the definition and characterization of aggregated contact networks [16,21,22] between individuals over arbitrary timescales, ranging from the finest time resolution of 20 seconds up to the entire duration of the event. Similarly to other studies on dynamic communication networks [16,22], this analysis is particularly insightful because it allows the acquisition of a system perspective where the statistical properties of individuals with similar interaction patterns can be identified. The aggregated network is defined as follows: nodes represent individuals, and an edge between two nodes represents an interaction that occurred between those nodes during the aggregation time interval. Each edge is weighted either by the total number of packets exchanged by the pair of tags, or by the total time during which the individuals have been in contact. We have verified that both definitions give the same results.

Figure 4 gives the main characteristics of the aggregated network for a time window of 12 hours during the *25C3* deployment, with the total number of exchanged packets used to define the intensity of the link (see Figures S5 and S6 for other examples). The distribution of weights is broad, showing that the heterogeneity in the duration of individual contacts (Figure 2) persists when contact durations are cumulated over a long time interval. The strength of a node is given by the sum of the weights of its links [23] and therefore represents, for each individual, the total time of interaction with other individuals. The corresponding distributions, reported in Figures S5 and S6, are also broad and display a large heterogeneity of behavior in the interaction patterns of individuals. Strikingly, the node strength grows super-linearly with the degree, i.e., the cumulated time of interaction of a given individual grows super-linearly with the number of distinct persons that this individual has had contacts with. This is a rather consistent observation across our experiments. In other words, the more distinct interactions one individual has and the larger is the average time dedicated to those interactions. This is in contrast with the sub-linear behavior that has been reported for mobile phone activity [16]. The super-linear association between the number of contacts and their average duration is the statistical signature of super-connectors that not only develop a large number of distinct interactions, but also dedicate an increasingly larger amount of time to such interactions. These highly social individuals are the crucial actors in defining the pattern of spreading phenomena [24-26]. The role of super-spreaders has been emphasized in the epidemiological and physics literature since a long time. However, the results emerging from our study indicate that super-spreaders may have a much larger spreading ability than what could be expected from just harnessing the number of their distinct contacts. The dynamical dimension provided by the high temporal resolution of the presented experiments might be the key to gather new data on the interplay between the concurrency/duration of contacts and their number [27]. We stress that the non-linear statistical association highlighted by our results is not a natural feature of most network models. Along with the possibility of considering generative models that reproduce this feature, it is worth considering the obtained datasets – and those that will be collected in future experiments – as test-beds for the investigation of diverse dynamical phenomena that take place on dynamical networks of human contact, such as social contagion and the propagation of airborne infections. The ability to resolve the least known scale of face-to-face presence for communities of several hundred persons is a critical enabler for these studies. Finally, our results may open the path to additional studies about the fundamental mechanisms of human interaction that underlie unexpected scaling behaviors observed at different levels of social aggregation [28-30].

In conclusion, this paper presents an experimental platform for gathering data on the social interactions of individuals that reconciles scale and detail through the use of low cost active RFID devices designed to operate as a distributed proximity sensing network. We present the results of three studies where the RFID platform was deployed in different contexts. Novel aspects of human dynamics and social interactions are found that highlight the emergence of structural and temporal features as a result of the inclusion of the dynamics in defining the structure of the network. At the micro-level, this experimental framework provides a new approach for the unsupervised collection of social interaction data, opening the path to the understanding and characterization of interaction mechanisms that represent the basic ingredients of realistic agent-based models for diseases and information spreading phenomena. In addition, the devices brings about the potential for attaining a multi-scale view of social interactions, while paving the way for a range of developments and applications.

## Materials and methods

The experiments we perform consist of a distributed sensing component, comprising wearable active RFID (Radio Frequency Identification) devices, and of a data collection and processing component comprising RFID readers installed in the environment, a local area network (LAN) and a central computer system that collects and stores the data. In the following we outline the recruitment process, the architecture and function of these components.

*Ethics Statement*
The recruitment and data collection were organized locally at each event. Attendees in the *SFHH* deployment and lab members in the *ISI* deployment were invited to participate by signing a written informed consent in conformity with the privacy regulations of the country laws where the experiment took place. The data have been collected in such a manner that subjects cannot be identified, directly or through identifiers linked to the subjects. Data collection, encryption, usage, and analysis were conducted in conformity with the EU regulations on privacy matters for scientific purposes, as detailed in the document for the informed consent. No data of ethical concern (personal information, medical records, etc.) have ever been collected. The *25C3* data collection has been organized by the OpenBeacon project and the raw data are publicly available (http://people.openbeacon.org/meri/openbeacon/sputnik/data/25c3).

*Distributed proximity sensing*

We use the exchange of low-power radio packets between wearable devices as a proxy for the spatial proximity of the individuals wearing such devices. The wearable device we use, shown in Figure S7, is an active RFID (Radio Frequency Identification) tag based on a design developed by the OpenBeacon project (http://www.openbeacon.org). The standard behavior of an active RFID tag is that of a radio beacon, i.e., at regular intervals of time the device emits a radio packet that carries a unique identificator associated with the device. The devices we use operate in the 2.4 GHz ISM band of the RF spectrum, and are based on the Nordic Semiconductor nRF24L01+ Single Chip Transceiver.

In the context of our experiments, we re-designed the RFID tags so that, in addition to their standard behavior, they also engage in bi-directional communication among themselves, in a peer-to-peer fashion. The devices perform a scan of their neighborhood by alternating transmit and receive cycles. During the transmit phase, low-power packets are sent out on a specific radio channel, hereby called the *contact channel*. During the receive phase, the devices listen on the same channel for packets sent by nearby devices. By including the transmit signal strength in the payload, the receiving device can estimate the degree of proximity of the transmitting device, and this operation can be carried out in a decentralized fashion throughout the sensing network. The lowest power level we use in our experiments is chosen so that packet exchange at that power level is only possible between devices situated within 1-1.5m of one another. Tags in close proximity exchange with one another a maximum of about 1 low-power packet per second.

In our experiments, active RFID tags are either secured to the lanyards that hold conference badges, clipped to the clothing of participants at the chest level, or inserted in the conference badge holders. In all cases, the antenna of the RFID tag is laying close to the skin of the participant, in the upper and frontal region of the body. The radio frequency emitted by the RFID tag is absorbed by body water. Because of this, the low-power packets we use for proximity sensing can only propagate towards the front of the individual wearing the device. At a fixed distance, this introduces an extremely strong anisotropy in the packet exchange rate that depends on the face-to-face orientation of the persons wearing the devices. Exchange of these low-power packets thus becomes a proxy for face-to-face proximity of individuals. The line of sight between two devices that can exchange radio packets lies in a solid angle that is narrow enough to generate very few false positives in crowded situations. This was verified for example by monitoring the contacts recorded in a crowded room during a conference session [31], a situation where a high density of individuals wearing the tags practically does not lead to the detection of contact pairs. The audience is indeed facing towards the speaker and face-to-face interactions are absent, except for situations in which neighbors may shortly interact for exchanges of comments. The situation readily changes when the session breaks and participants start to interact and contact pairs are detected.

The rate at which low-power packets are emitted and the fraction of time the devices spend listening on the contact channel are tuned so that the face-to-face proximity of two individuals wearing the RFID tags can be assessed with a probability in excess of 99% over an interval of 20 seconds. This sets the natural time scale over which we perform the temporal aggregation of data collected from different devices.
Figure 1A summarizes the proximity detection strategy: if two individuals are not within 1-1.5m of each other, no packet exchange is possible at the lowest power level used in the contact channel. The same is true is the individuals are nearby but are not facing each other. When two individuals are nearby and facing each other, low-power packet exchange occurs in either direction (1), is detected and reported to the data collection infrastructure (2) on a different radio channel, hereby called the *infrastructure channel*.

*Data collection infrastructure*
The spatial proximity relations are relayed from RFID tags to radio receivers, called RFID readers, installed in the experimental area. The radio receivers are connected to a central computer system by means of a Local Area Network. The readers listen on the infrastructure radio channel for incoming packets, and whenever they receive a packet they encapsulate it in a UDP (User Datagram Protocol) packet and relay it to a central server, where it is timestamped and stored.
The received packets are also fed to a real-time system that aggregates them and maintains a real-time graph representation of the proximity relations among experiment participants. This representation is used for analysis, for visualization (see below) as well as to run user-oriented applications.

The top panel of Figure 5 illustrates how the global contact graph among individuals is built, at a given time, by aggregating the proximity information reported by single devices over a sliding window of $\Delta t = 20$ seconds. This instantaneous contact graph is represented as a time-dependent adjacency matrix $A^t_{ij}$, such that $A^t_{ij} = 1$ if the RFID tags i and j exchanged at least one packet at the lowest radio power during the time inteval $[t-\Delta t, t]$, and $A^t_{ij} = 0$ otherwise. This network representation of the face-to-face proximity relations, computed as a function of time for an entire experiment, is the basic piece of information that we use for the analysis.

More details about the distributed proximity-sensing system we developed are available on the web site of the SocioPatterns project, http://www.sociopatterns.org.

*Visualization*

The deployments we conduct are accompanied with publicly displayed dynamic visualizations of the proximity relations between individuals. Two types of visualizations are displayed. The first is a dynamic representation of the *instantaneous* network of proximity. The second represents the *cumulative* network of contacts, which summarizes the amount of time each pair of individuals spent together, as measured from the beginning of the experiment.

A snapshot of the real-time visualization is shown in the bottom panel of Figure 5. A force-directed graph layout algorithm is used to display the current state of the network. The proximity graph is computed in real-time by the data collection system, and the visualization is updated continuously [32].

## Acknowledgments

We thank Bitmanufaktur, the OpenBeacon project, the organizers of the SFHH conference, the SFHH and GOJO. CC acknowledges technical support from Milosch Meriac and Brita Meriac. The authors acknowledge the valuable feedback, patience and support of the over 800 volunteers who participated in the deployments.

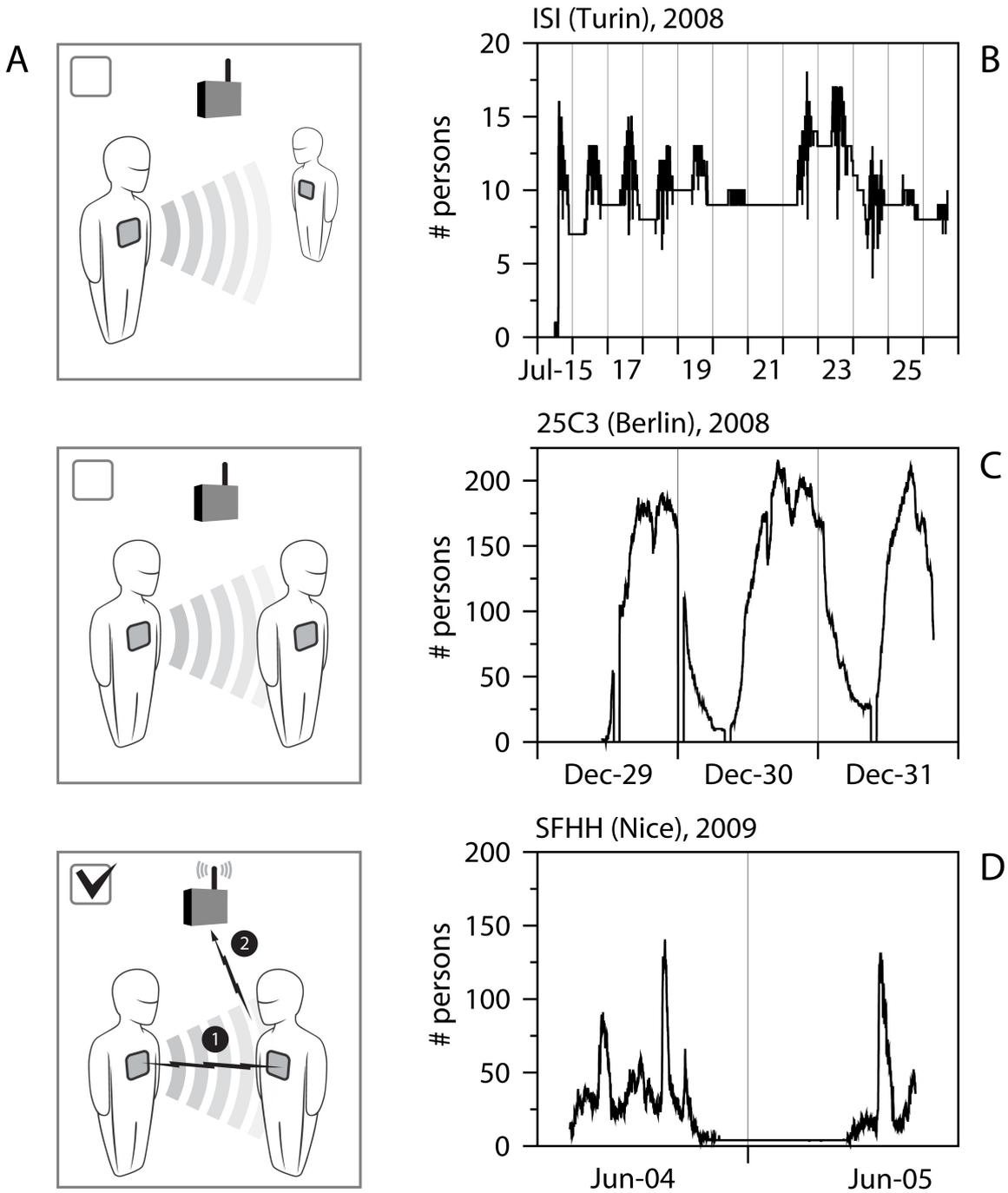

**Figure 1.  RFID sensor system and system deployments.** A) Schematic illustration of the RFID sensor system. RFID tags are worn as badges by the individuals participating to the deployments. A face-to-face contact is detected when two persons are close and facing each other. The interaction signal is then sent to the antenna. B)C)D) Activity pattern measured in terms of the number of tagged individuals as a function of time in the three deployments: B) *ISI* refers to the deployment in the offices of the ISI foundation in Turin, Italy, with 25 participants; C) *25C3* to the 25th Chaos Communication Congress in Berlin, Germany, with 575 participants, and D) *SFHH* to the congress of the Société Française d'Hygiène

# Dynamics of person to person interactions

Hospitalière, Nice, France, with 405 participants. Dashed vertical lines indicate the beginning and end of each day. Typical daily rhythms are observed in the office and conference settings. The *ISI* deployment allows us to recover the weekly pattern signaled by the absence of activity on the day of Sunday (the number of persons larger than zero at night indicates the tags left in the offices, easily recognizable from the flat behavior).

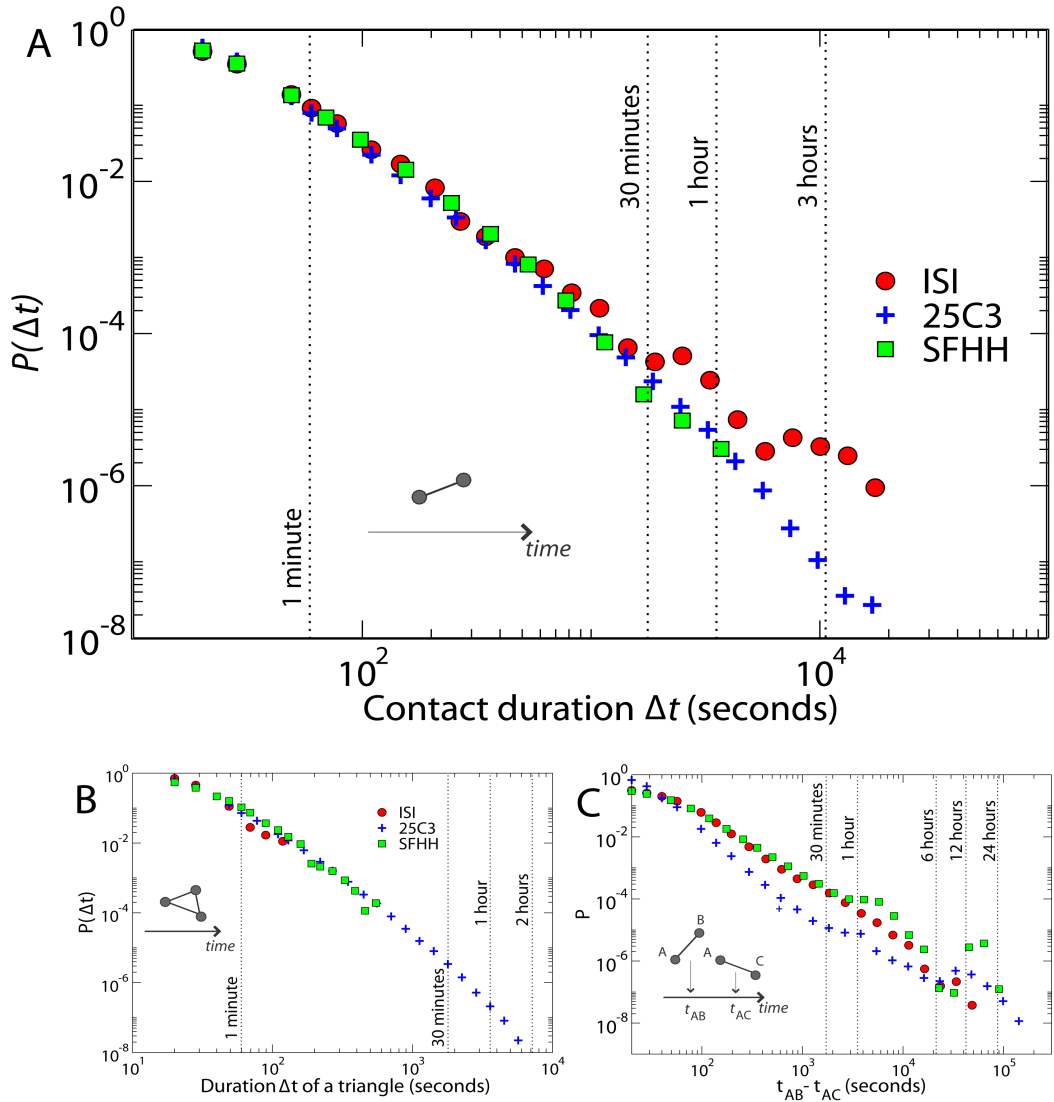

**Figure 2. Probability distribution of human interactions.** A) Probability distribution of duration of contacts between any two given persons. Strikingly, the distributions show a similar long-tail behavior independently of the setting or context where the experiment took place or the detection range considered. The data correspond to respectively 8700, 17000 and 600000 contact events registered at the *ISI*, *SFHH* and *25C3* deployments. B) Probability distribution of the duration of a triangle. The number of triangles registered are 89, 1700 and 600000 for the *ISI*, *SFHH* and *25C3* deployments. C) Probability distribution of the time intervals between the beginning of consecutive contacts *AB* and *AC*. Some distributions show

spikes (i.e., characteristic timescales) in addition to the broad tail; for instance, the 1h spike in the *25C3* data may be related to a time structure to fix appointments for discussions.

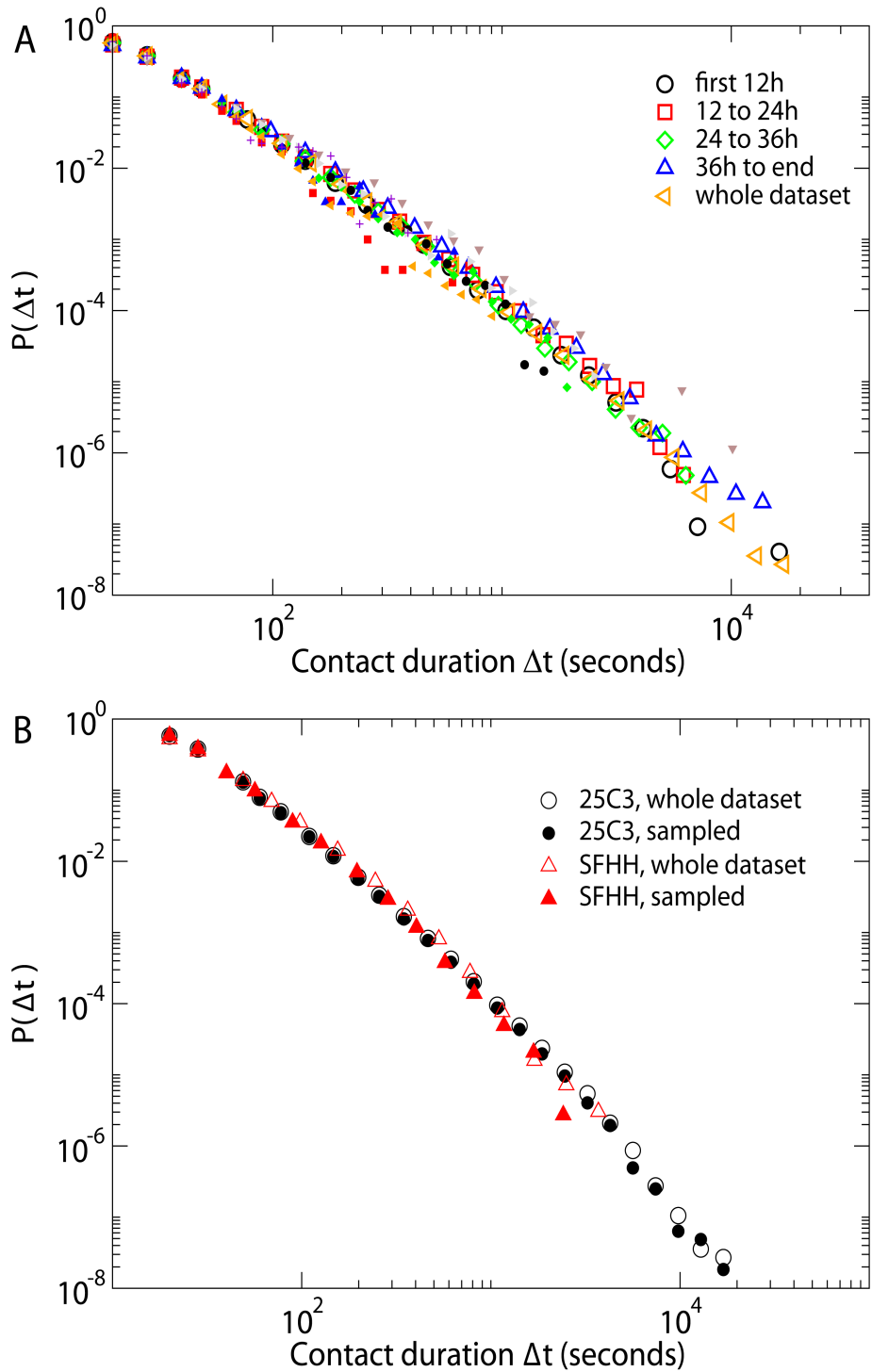

**Figure 3. Robustness.** A) Distribution of contact durations (in seconds) at the *25C3* deployment, for various time intervals and for the entire dataset. The filled symbols correspond to the distribution of contact durations of several individual

tags. B) Distribution of contact durations (in seconds) for sampled datasets in which 60% of the tags are ignored, compared with the distributions obtained from the whole datasets.

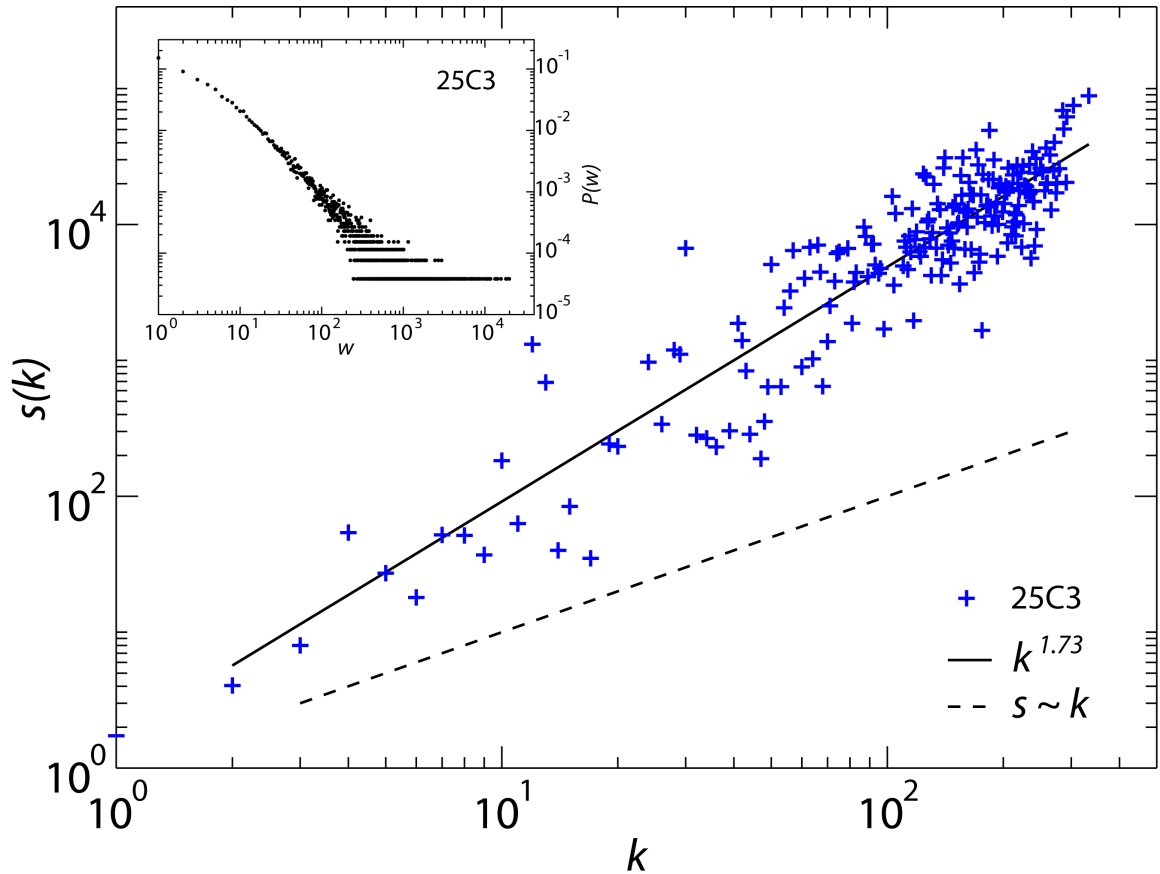

**Figure 4. Network properties.** Properties of the aggregated network of contacts corresponding to the third 12-hour period of the 25C3 deployment. The total number of packets exchanged by a tag during a contact (strength $s$) is shown as a function of the number of distinct contacts (degree $k$). A superlinear (powerlaw) behavior is observed, with a slope of 1.73 [95%CI: 1.65-1.81] obtained from the fitting procedure with a correlation coefficient of 0.93. Inset: distribution of links' weights, defined as the total number of packets exchanged between two interacting tags. The same qualitative properties are obtained for other time intervals and for all the other experiments we deployed.

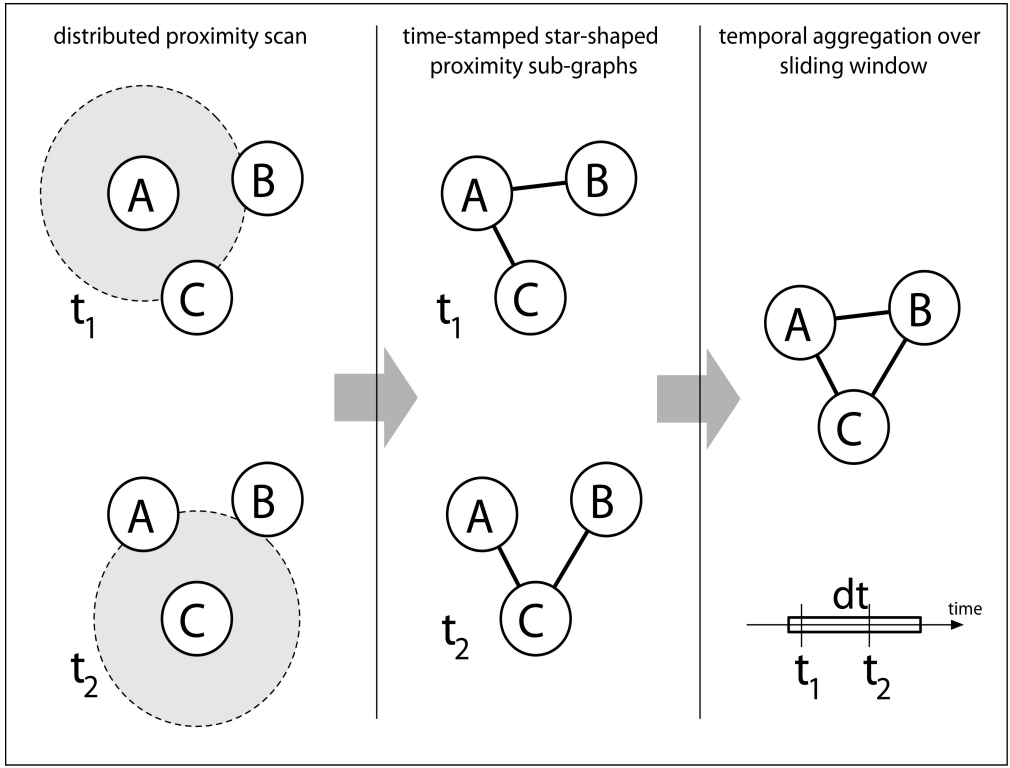

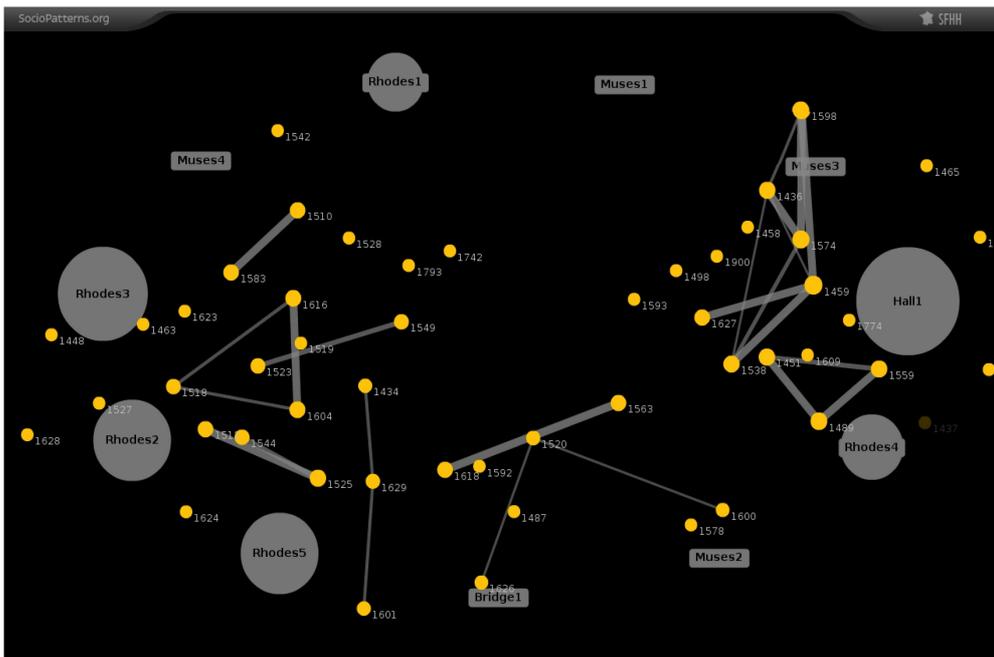

**Figure 5. From RFID communications to contact networks.** Top: Temporal aggregation of proximity relations reported by different tags over a sliding window. The information collected by each tag is aggregated and translated into a dynamical adjacency matrix to reconstruct the dynamical network of face-to-face interactions. Bottom: Real-time visualization. A snapshot of the visualization, displaying approximate position information as well as the instantaneous network of face-to-face proximity. Individuals wearing an RFID tag are represented as discs labeled with the numeric identifier of their tag. Edges between individuals represent ongoing

face-to-face proximity relations, and their thickness reflects the strength of the proximity relations. The other labels refer to names of rooms in the venue and denote the location of RFID readers. The graph is laid out so that individuals are shown near the readers that report their presence, and the sizes of the readers symbols depend on the number of users from which they receive information.

## Supplementary Information

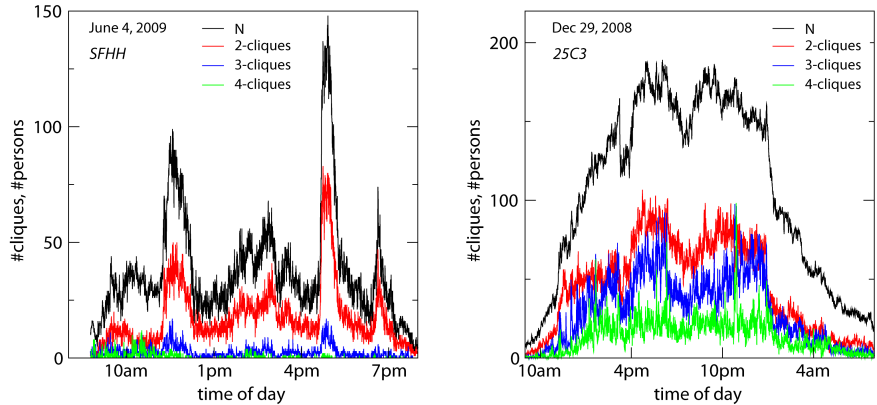

**Figure S1.** Activity timeline for the first day of the *SFHH* deployment (left) and for the second day of the *25C3* deployment (right). The figures show the number of tags (black), the number of pairs (red), triangles (blue), and 4-cliques (green) in the contact network aggregated over a sliding window of 20 seconds, as a function of time.

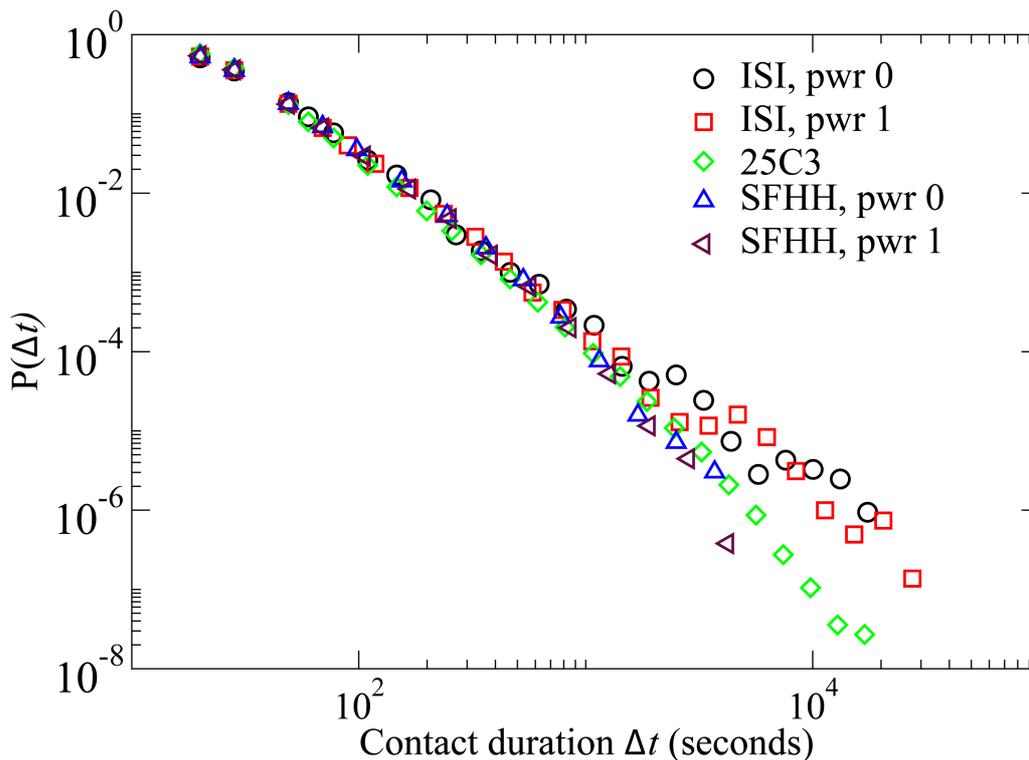

**Figure S2.**
Distribution of contact durations (in seconds) for all experiments performed and for the different available detection ranges.

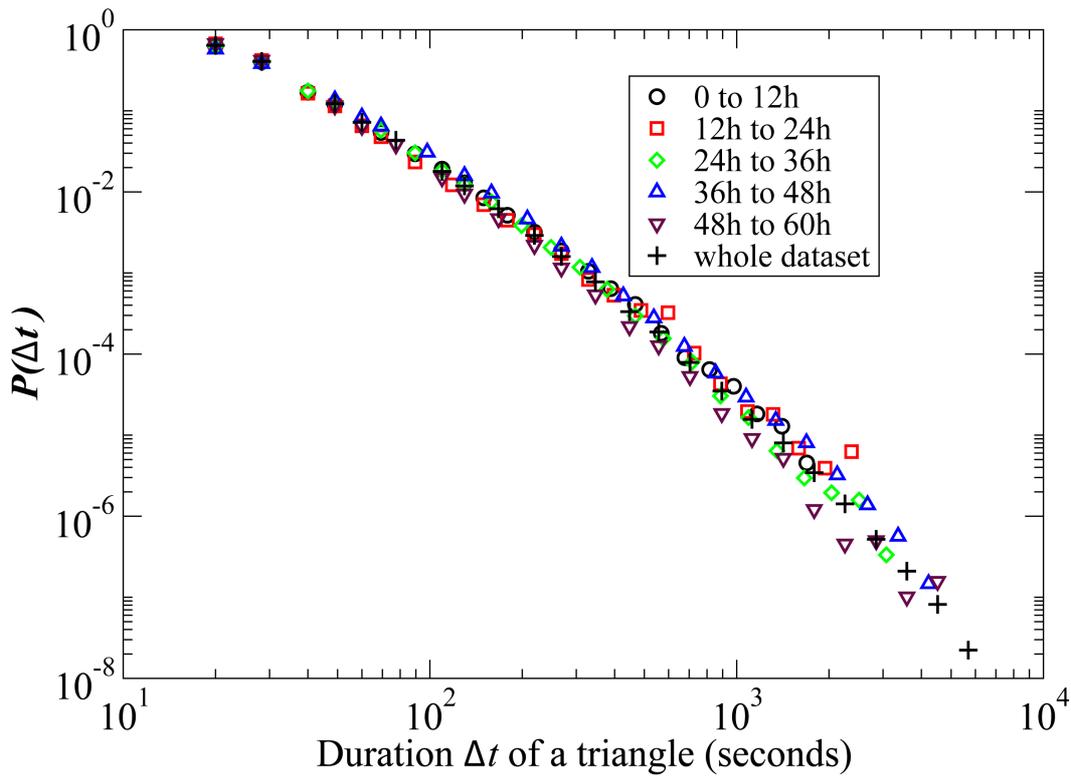

**Figure S3.** Distribution of triangle durations (in seconds) at the *25C3* deployment, for several time intervals.

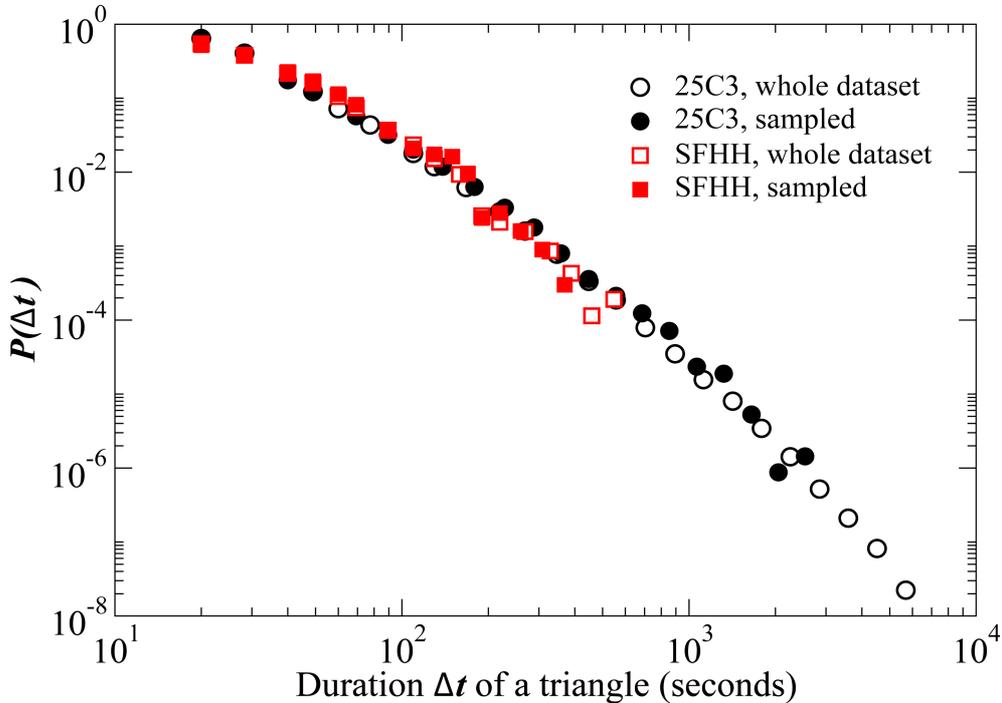

**Figure S4.** Distribution of triangle durations (in seconds) for sampled datasets in which 30 to 60% of the tags are ignored, compared with the distributions obtained

from the whole datasets.

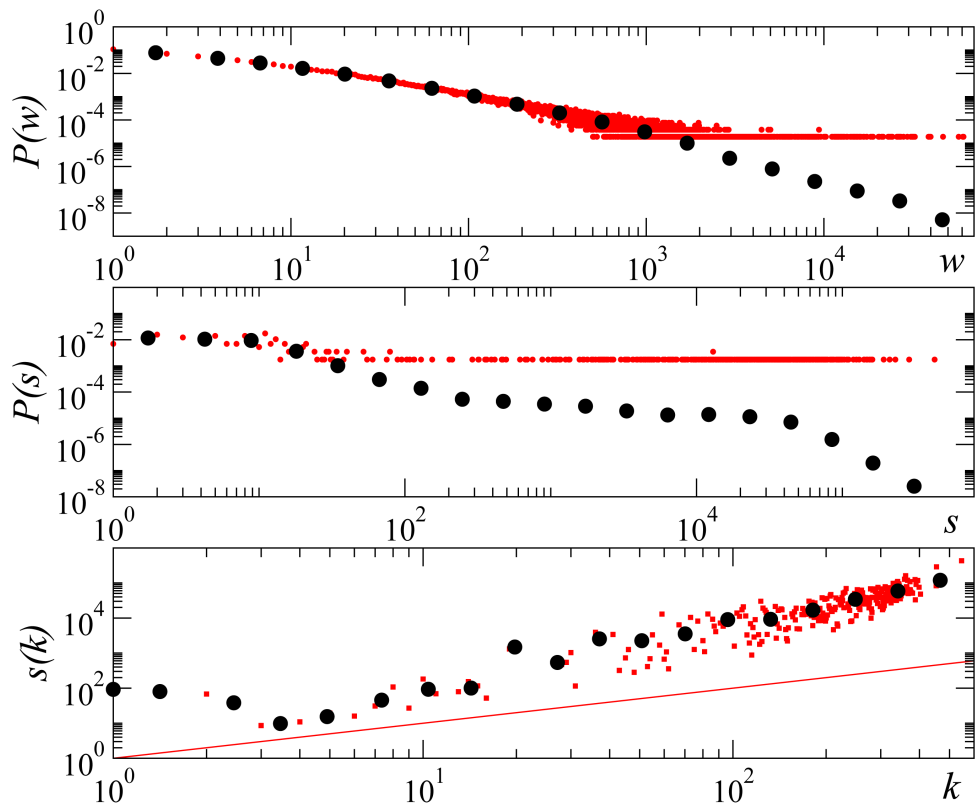

**Figure S5.**
Some characteristics of the aggregated network of contacts corresponding to the whole *25C3* deployment. From top to bottom: distribution of the edge weights, of node strengths, and node strength as a function of node degree. Red dots display the raw data, and black circles are log-binned data. The red line shows a linear behavior s ~k.

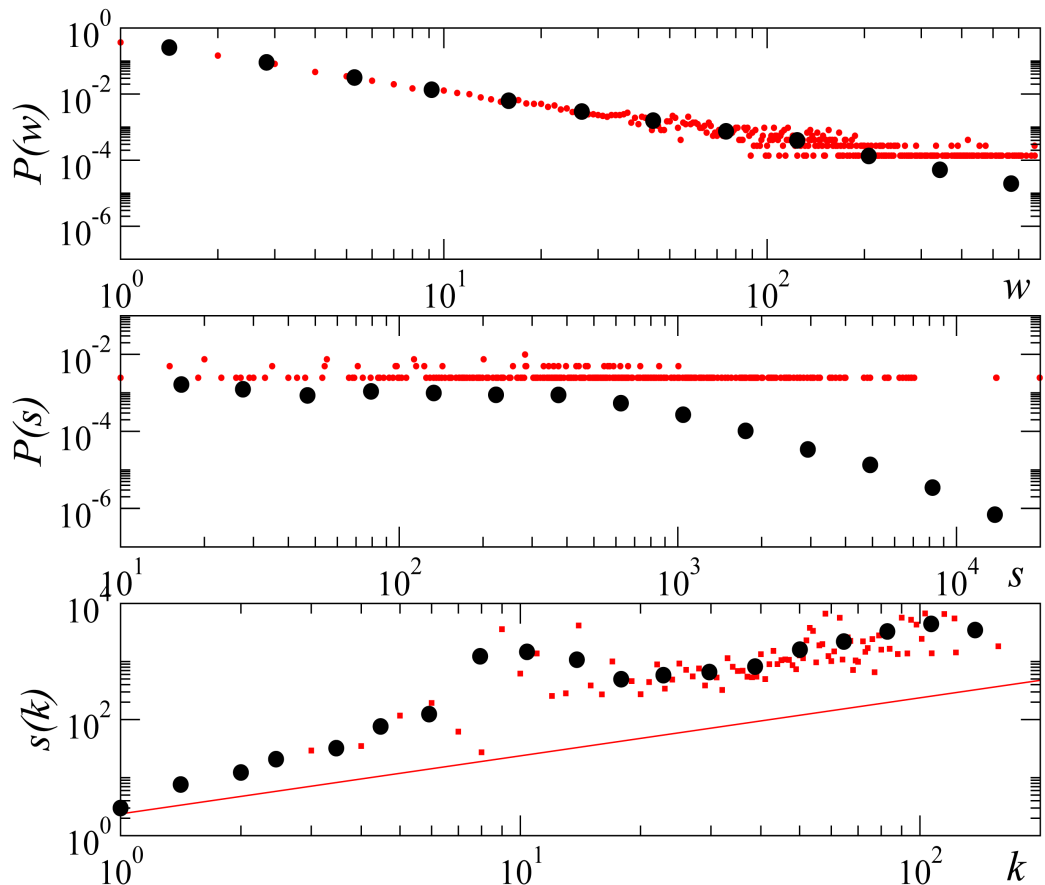

**Figure S6.** Same as Fig. S5 for the network of contacts aggregated of the *SFHH* deployment.

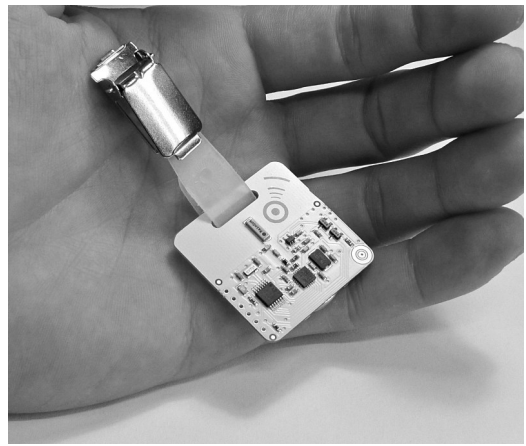

**Figure S7.** Active RFID tag used in the experiments. The RFID tag is based on an open design by the OpenBeacon project and features a microcontroller, a radio transceiver operating in the 2.4 GHz ISM band, an antenna embedded in the printed circuit board, and a lithium battery.